\documentclass{aa}

\usepackage{graphicx,natbib,txfonts}
\bibpunct{(}{)}{;}{a}{}{,} 

\begin{document} 

\title{Lyot-plane phase masks for improved high-contrast imaging\\with a vortex coronagraph}

\author{G. J. Ruane\inst{\ref{inst1},\ref{inst2}}\and
		E. Huby\inst{\ref{inst1}}\and
		O. Absil\inst{\ref{inst1}}\fnmsep\thanks{F.R.S.-FNRS Research Associate}\and
		D. Mawet\inst{\ref{inst3}}\and
		C. Delacroix\inst{\ref{inst4}}\and
		B. Carlomagno\inst{\ref{inst1}}\and
		G. A. Swartzlander, Jr.\inst{\ref{inst2}}
}

\institute{D\'{e}partement d'Astrophysique, G\'{e}ophysique et Oc\'{e}anographie, Universit\'{e} de Li\`{e}ge, All\'{e}e du Six Ao\^{u}t 17, B-4000 Li\`{e}ge, Belgium\label{inst1}
         \and
Chester F. Carlson Center for Imaging Science, Rochester Institute of Technology, 54 Lomb Mem. Dr., Rochester, NY 14623, USA\label{inst2}
         \and
California Institute of Technology, 1200 E. California Blvd., Pasadena, CA 91125, USA\label{inst3}
         \and
CRAL, Observatoire de Lyon, CNRS UMR 5574, Universit\'{e} Lyon 1, 9 avenue Charles Andr\'{e}e, 69230 Saint-Genis Laval, France\label{inst4}
}


\abstract
{The vortex coronagraph is an optical instrument that precisely removes on-axis starlight allowing for high contrast imaging at small angular separation from the star, thereby providing a crucial capability for direct detection and characterization of exoplanets and circumstellar disks. Telescopes with aperture obstructions, such as secondary mirrors and spider support structures, require advanced coronagraph designs to provide adequate starlight suppression.}
{We introduce a phase-only Lyot-plane optic to the vortex coronagraph that offers improved contrast performance on telescopes with complicated apertures. Potential solutions for the European Extremely Large Telescope (E-ELT) are described and compared.}
{Adding a Lyot-plane phase mask relocates residual starlight away from a region of the image plane thereby reducing stellar noise and improving sensitivity to off-axis companions. The phase mask is calculated using an iterative phase retrieval algorithm.}
{Numerically, we achieve a contrast on the order of $10^{-6}$ for a companion with angular displacement as small as 4~$\lambda/D$ with an E-ELT type aperture. Even in the presence of aberrations, improved performance is expected compared to either a conventional vortex coronagraph or optimized pupil plane phase element alone.}
{}


\maketitle

\section{Introduction}
The vortex coronagraph (VC) is an optical system for high-contrast imaging of astronomical objects at small angular separations \citep{Mawet2005,Foo2005}. The VC suppresses the light from a star allowing direct detection of dim companions, exoplanets, and circumstellar disks. Imaging objects with a VC that are otherwise buried in the noise associated with the bright host star has been demonstrated in laboratory \citep[e.g.][]{Mawet2009,Delacroix2013} and on-sky observations \citep[e.g.][]{Swartzlander2008,Mawet2010,Serabyn2010,Absil2013,Defrere2014}.

The VC was originally devised for telescopes with a circular aperture, where light from an on-axis point source is completely rejected while light from off-axis sources is preserved \citep{Mawet2005,Foo2005}. Unfortunately, most telescope pupils have an obstructing secondary mirror with spider support structures, for which the on-axis source is only partially suppressed \citep[e.g.][]{Jenkins2008}. Moreover, very large apertures are often formed by a segmented mirror, which is generally not circular and may have discontinuities between segments. 

Several solutions have been proposed to improve contrast performance of a VC on telescopes with complicated apertures. The effect of the secondary mirror may be mitigated by use of a sub-aperture \citep{Mawet2010,Ruane2013}, tandem coronagraphic stages \citep{Mawet2011_improved, Galicher2011, Mawet2013_spie}, or a ring-shaped apodizer \citep{Mawet2013_ringapod}. Spiders and other aperture discontinuities may be compensated for by binary amplitude apodizers \citep{Carlotti2014}, focal plane phase corrections \citep{Ruane2015_SPIE,Ruane2015}, or a pair of deformable mirrors \citep{Pueyo2013}.

Whereas a VC uses a focal-plane phase mask and downstream aperture stop ("Lyot stop") to suppress starlight, an alternate class of pupil-only coronagraphs have achieved considerable success on telescopes with complicated apertures, using amplitude \citep{Kasdin2003,Carlotti2011} and phase \citep{Codona2004,Kenworthy2007,Kenworthy2010} pupil elements. 

Here, we present continuous pupil-plane phase elements that introduce a spatially variant phase shift in the plane of the Lyot stop (see Fig. \ref{fig:VCschematic}), which acts to relocate residual starlight away from a defined region of interest where dim sources may be detected. The Lyot-plane phase mask (LPM) is intrinsically lossless, may be designed for a large variety of apertures, and is simple to integrate with conventional coronagraph designs. We show that improved contrast performance may be achieved compared to pupil-only phase mask coronagraphs. 

\section{The vortex coronagraph}

\begin{figure}[t]
	\centering
	\includegraphics[width=\linewidth,trim = 0 1.5mm 0 1.5mm,clip=true]{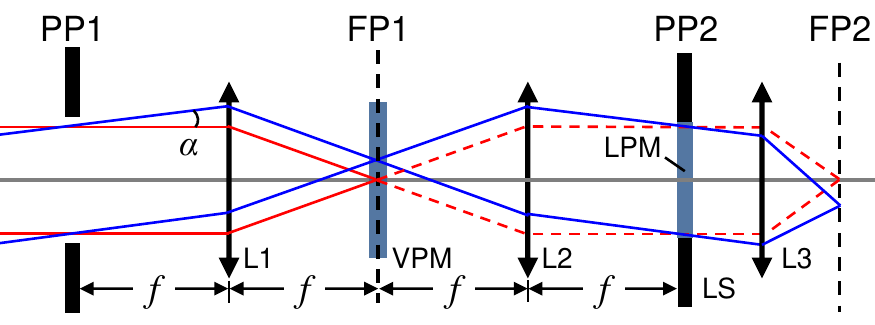}
	\caption{Schematic of a vortex coronagraph. The first pupil plane (PP1) is typically an image of the telescope aperture. Lens L1 forms focal plane FP1 with field $F(\rho,\phi)$. A phase-only element known as a vortex phase mask (VPM) is located at FP1, with transmission $t=\exp(il\phi)$, where $l$ is a nonzero, even integer. Lens L2 forms the output pupil plane (PP2) with field $E(r,\theta)$. The on-axis starlight ($\alpha=0$, red rays) is diffracted outside of the Lyot stop (LS). Lens L3 forms the subsequent focal plane (FP2) with the on-axis starlight removed, while light from off-axis sources ($\alpha\ne0$, blue rays) propagates through the LS. A Lyot-plane phase mask (LPM) may be introduced at PP2 to improve the contrast between off-axis and on-axis sources in the image at FP2. Lenses L1 and L2 have focal length $f$.}
	\label{fig:VCschematic}
\end{figure}

In this section, we demonstrate that the contrast performance of a conventional VC is limited on telescopes with complicated apertures. But first, we briefly review the case where the pupil is circular and has no obstructions. The layout of the VC is illustrated in Fig. \ref{fig:VCschematic}. A vortex phase mask (VPM) is placed in the focal plane (FP1) of a 4-$f$ optical system with transmission $t=\exp(il\phi),$ where $\phi$ is the azimuthal angle in FP1 and $l$ is an integer known as the "topological charge." A nonzero, even value of $l$ is required for ideal starlight suppression with a circular aperture. Various techniques are available to fabricate such achromatic phase masks \citep{Bomzon2001,Marrucci2006,Murakami2012}. The scalar field immediately after the VPM owing to an on-axis point source may be written
\begin{equation}
F\left(\rho,\phi \right)=\frac{k a^2}{f}\frac{J_1\left( k a \rho/f\right)}{k a \rho/f} e^{il\phi},
\label{eq:PSF}
\end{equation}
where $\rho$ is the radial coordinate in FP1, $a$ is the pupil radius, $k = 2\pi/\lambda,$ $\lambda$ is the wavelength, $f$ is the focal length, and $J_n$ is the $n$th order Bessel function of the first kind. The field at the output pupil (PP2) is given by the Fourier transform of Eq. \ref{eq:PSF}:
\begin{equation}
E\left(r,\theta \right)=e^{i2\theta}\frac{ka}{f}\int\limits_0^\infty{J_1\left(k a \rho/f\right)J_l\left(k r \rho/f\right)d\rho},
\label{eq:pp2integral}
\end{equation}
where $\left(r,\theta \right)$ are the polar coordinates in PP2. Remarkably, Eq.~\ref{eq:pp2integral} evaluates to a discontinuous function, which is zero-valued within the geometric image of the pupil ($r<a$) for nonzero, even values of $l$; that is, a nodal area appears. All of the light from a distant on-axis point source appears outside of this region; for example, $E\left(r,\theta \right)=(a/r)^2e^{i2\theta}$ for $r>a$ and $l=2$. A circular aperture stop known as the Lyot stop (LS), with radius $a_L$ where $a_L \le a$, is placed in PP2 to block all of the light from the on-axis source. Off-axis sources (i.e. $\alpha \ne 0$) do not form a nodal area and therefore transmit through the LS. For $\alpha \ll \lambda/D$, the transmitted energy increases as $\alpha^{|l|}$.

\begin{figure}[t]
	\centering
	\setlength{\fboxsep}{4pt}
	\setlength{\fboxrule}{0pt}
	\includegraphics[width=0.9\linewidth,trim = 0 -1mm 0 0,clip=true]{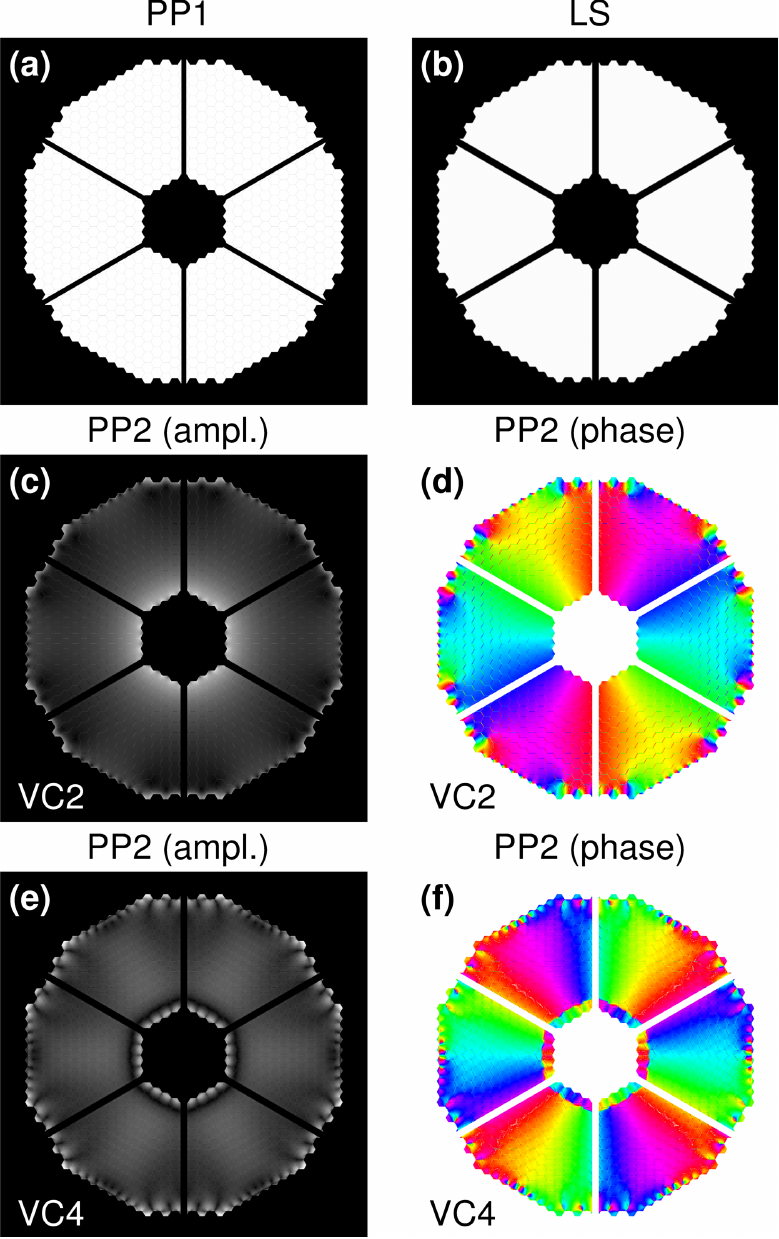}\\
	\includegraphics[trim = 0 0 -2mm 0,clip=true]{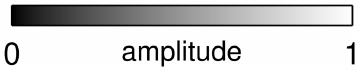}
	\includegraphics[trim = 0mm 0 0mm 0,clip=true]{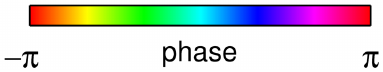}\\
	\caption{(a)~Pupil of the E-ELT. (b)~Lyot stop designed for a VC on E-ELT. (c)~Amplitude and (d) phase of the field directly after the LS for an on-axis point source and $l=2$. (e)-(f)~Same as (c)-(d), but for $l=4$.}
	\label{fig:PP1_PP2}
\end{figure}

Although the VC theoretically provides ideal starlight suppression with an unobstructed circular pupil, telescope apertures are often more complicated. For this discussion, we consider the future European Extremely Large Telescope (E-ELT). Figure \ref{fig:PP1_PP2} shows the entrance pupil of the E-ELT (see Fig. \ref{fig:PP1_PP2}(a)) and a possible LS (see Fig. \ref{fig:PP1_PP2}(b)) along with the complex fields at PP2 directly after the LS owing to an on-axis point source and a VC with $l=2$ (VC2, see Fig. \ref{fig:PP1_PP2}(c)-(d)) and $l=4$ (VC4, see Fig. \ref{fig:PP1_PP2}(e)-(f)). The fields here, and throughout this work, are calculated using the Fast Fourier Transform (FFT) algorithm on large computational arrays ($16384\times16384$ samples), with 1046 samples across the pupil diameter $D$ and 15.7 samples per $\lambda F\#$ in the image plane, where $F\#=f/D$. It can be seen in Figs. \ref{fig:PP1_PP2}(c) and \ref{fig:PP1_PP2}(e) that light from the on-axis point source leaks through the LS. We note that the LS is slightly smaller than the entrance pupil with a dilated central obscuration and spiders reducing the maximum system transmission (i.e. without a focal plane element) by 4\%. 

\begin{figure}[t]
	\centering
	\includegraphics[width=\linewidth,trim = 0mm 1.5mm 1mm 1mm,clip=true]{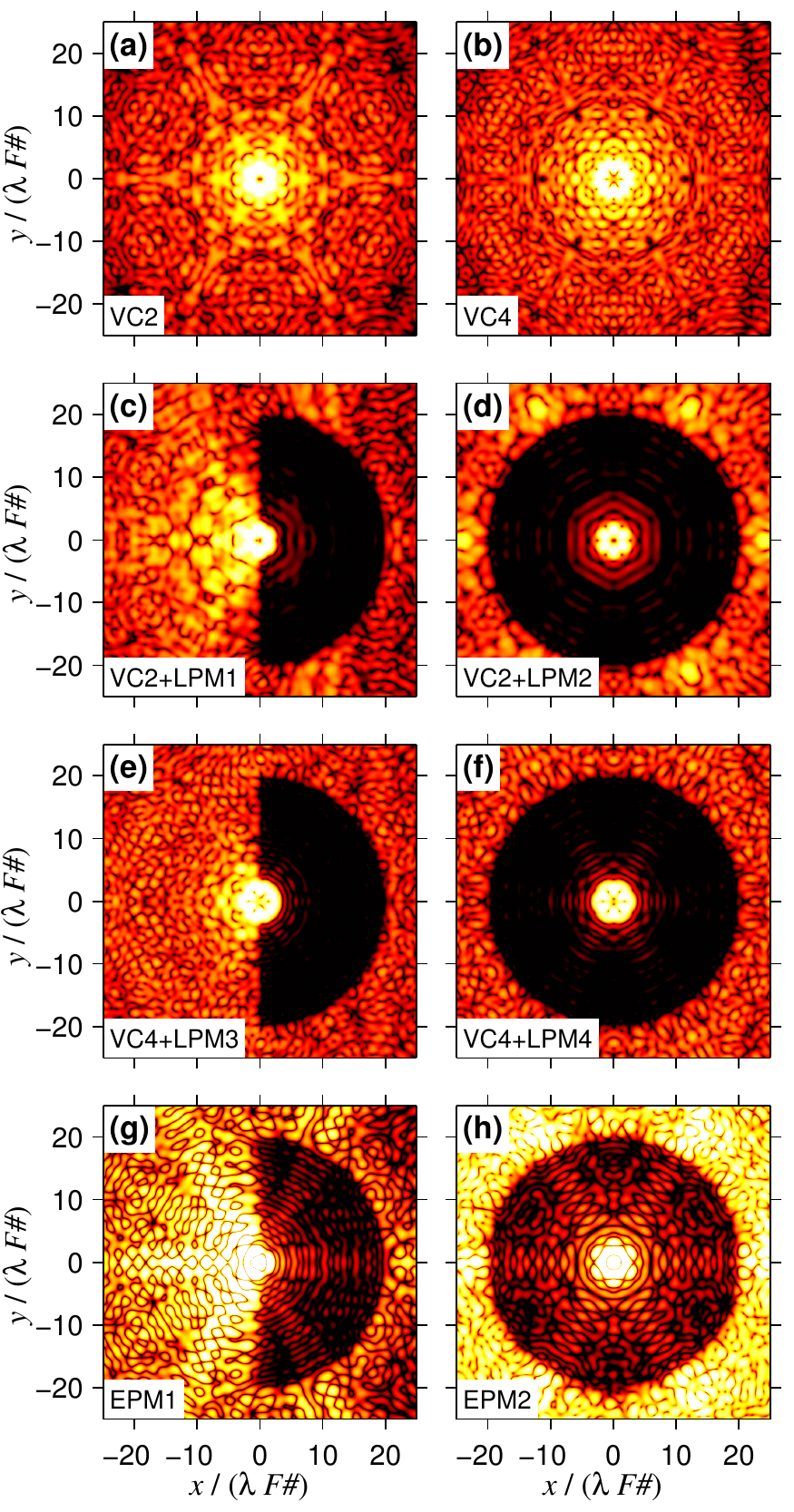}
	\includegraphics[trim = -7.5mm 0 0 0,clip=true]{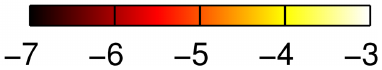}\\
	\caption{Monochromatic on-axis PSFs (in log irradiance) for (a)-(b)~VCs, (c)-(f)~VCs with LPMs, and (g)-(h) EPMs only. Examples phase masks that form semi-annular and full-annular dark regions are shown. Each is normalized by the peak value of the E-ELT PSF.}
	\label{fig:psfs}
\end{figure}

\begin{figure}[t]
	\centering
	\includegraphics[width=0.95\linewidth,trim = 0 0 0 1mm,clip=true]{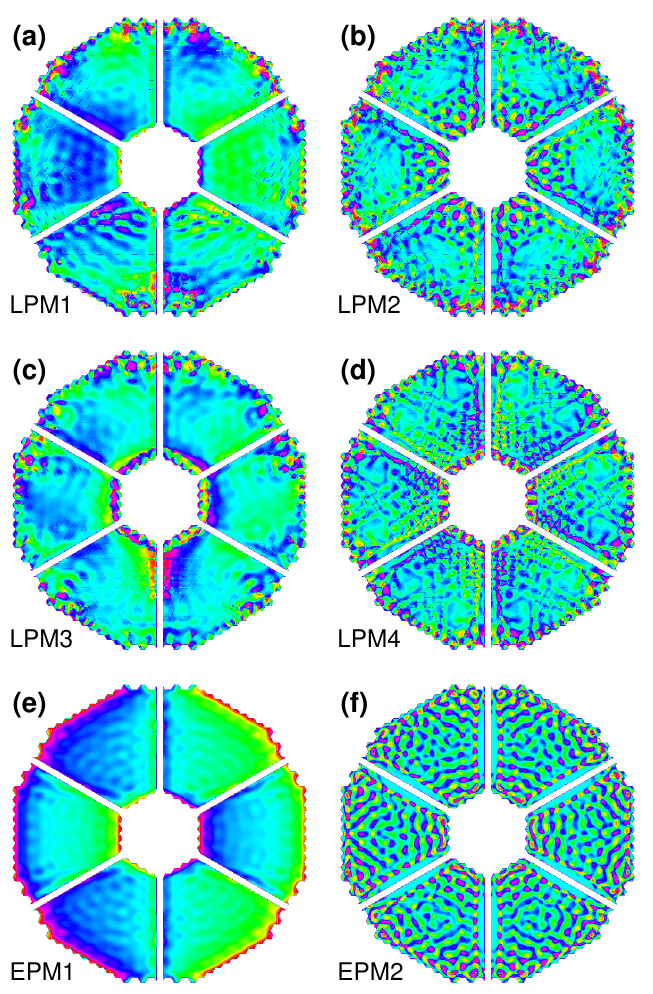}
	\includegraphics[trim = 0 0 -2mm 0,clip=true]{horizcolorbar_phase_hsv.pdf}\\
	\caption{Optimized pupil plane phase masks for (a)-(b)~VC2, (c)-(d)~VC4, and (e)-(f) no VPM. The corresponding PSFs are shown in Fig. \ref{fig:psfs}.}
	\label{fig:LPMs}
\end{figure}

\section{Lyot-plane phase masks for E-ELT}
In the case of the E-ELT pupil, the VPM and LS (together, the VC) block 95.5\% of the total power due to the star with $l=2$ and 95.3\% with $l=4$. The residual starlight reaches FP2 where it is spread over several units of $\lambda F\#$, along with its associated noise (see Fig. \ref{fig:psfs}(a)-(b)). Improved sensitivity in a given discovery region may be achieved by designing a dark hole in the point spread function (PSF) of an on-axis source (see Fig. \ref{fig:psfs}(c)-(h)). This approach has found success in the apodizing phase plate (APP) coronagraph \citep{Codona2004,Kenworthy2007,Kenworthy2010}. In comparison to a focal plane coronagraph, the APP uses a pupil-plane phase mask to produce a dark hole in the spatially invariant PSF rather than reducing the amount of on-axis starlight reaching the image plane. Our approach combines focal plane and pupil plane coronagraphy to first suppress the star with a VC and then sculpt a dark hole in the residual starlight with an LPM, further improving the contrast locally. 

To reduce unwanted starlight, we present two LPM designs for VC2 (LPM1 and LPM2, see Fig. \ref{fig:LPMs}(a)-(b)) and VC4 (LPM3 and LPM4, see Fig. \ref{fig:LPMs}(c)-(d)). LPM1 and LPM3 form a semi-annular dark hole in FP2 (see Fig. \ref{fig:psfs}(c),(e)), while LPM2 and LPM4 clear the full annulus (see Fig. \ref{fig:psfs}(d),(f)). We also present entrance pupil masks (EPM) that form semi and full annular dark holes (respectively EPM1 and EPM2), without the use of a VPM in FP1 (see Fig. 3(g)-(h) and Fig. 4 (e)-(f)). The EPM is similar in principle to an APP coronagraph \citep{Codona2004}. 

The LPMs and EPMs are calculated using a point-by-point iterative phase retrieval method \citep{Ruane2015_SPIE,Ruane2015}, where the star is assumed to be an infinitely distant point source and the field in PP2 is found computationally using paraxial Fourier optics methods. The phase in PP2 is optimized such that a dark hole appears over a pre-defined region in FP2. Then, the resulting phase mask needed to provide the necessary phase shift to match the optimized PP2 phase is calculated. We note that, for simplicity, all of the simulated optical configurations include the Lyot stop, even when there is no VPM or pupil plane mask present.

For exoplanet imaging, a dark hole in the on-axis PSF in FP2 is desired starting at a few $\lambda/D$ in angular separation from the star location to the edge of the field of view. Here, we chose to create a dark hole ranging from about $4~\lambda/D$ to $20~\lambda/D$. The outer boundary of the dark hole roughly corresponds to the control radius of state-of-the-art extreme adaptive optics systems like GPI \citep{Macintosh2014} and SPHERE \citep{Fusco2014}. 

\begin{figure*}[t]
	\centering
	\includegraphics[width=0.48\linewidth]{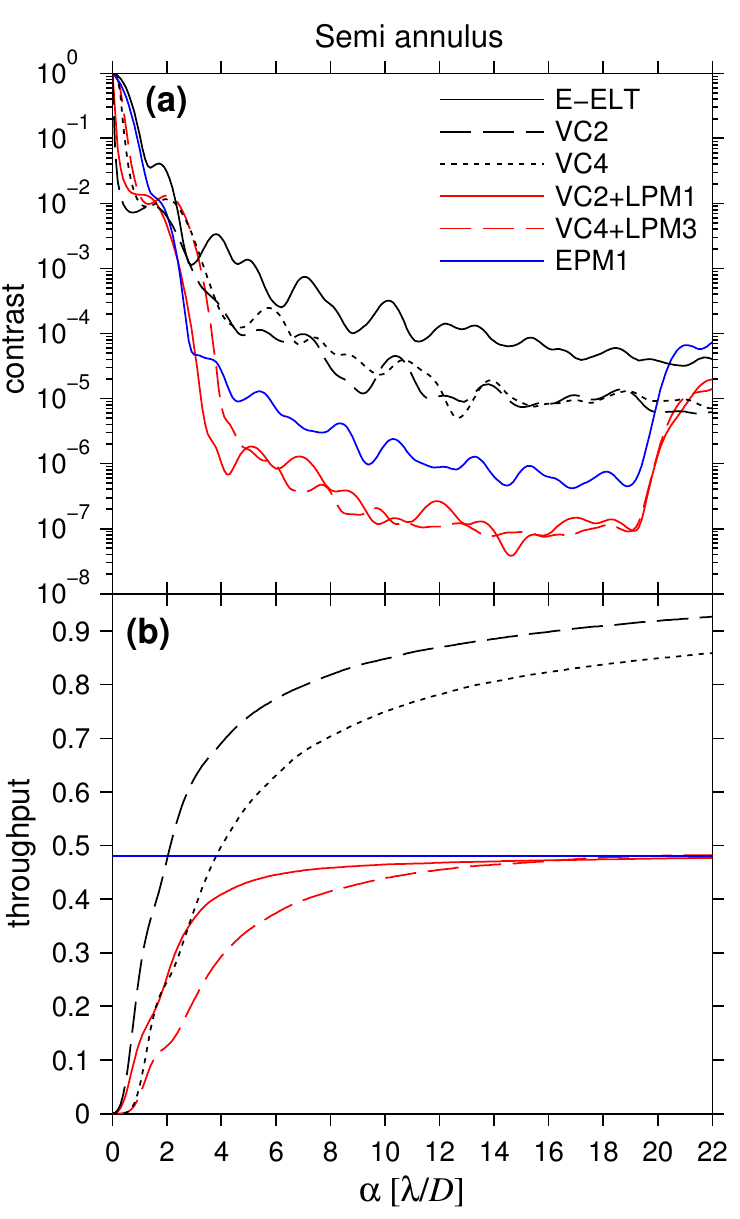}
	\includegraphics[width=0.48\linewidth]{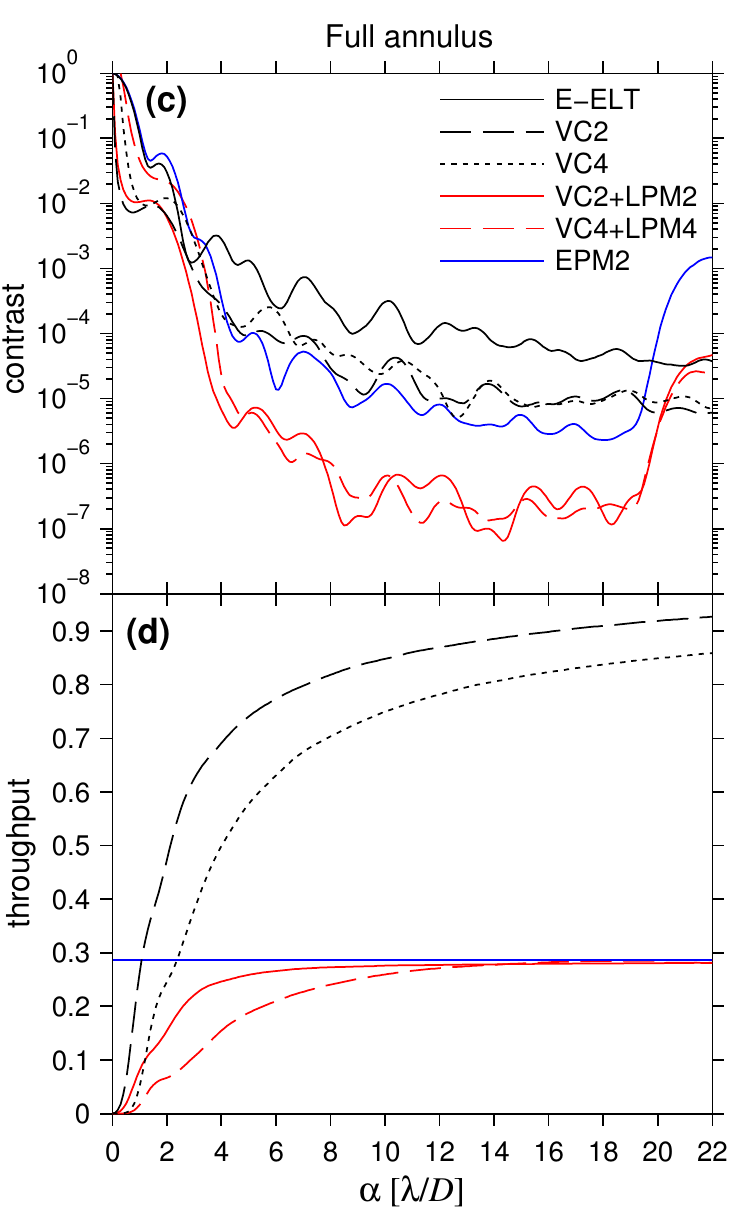}
	\caption{(a) Contrast and (b) throughput of an off-axis source as a function of angular displacement $\alpha$ for the LPMs optimized to produce a semi-annular dark region. (c)-(d) Same as (a)-(b), but for a full annulus. The contrast and throughput afforded by the E-ELT with and without EPMs are also shown for comparison.}
	\label{fig:offaxisperf}
\end{figure*}

\section{Performance}
During the phase optimization routine, the amount of light that appears in the dark region of the on-axis PSF in FP2 decreases iteratively. However, the off-axis PSF quality is also affected. We find the optimal phase mask by monitoring the contrast and throughput at each iteration. The former is defined as the energy ratio between on-axis and off-axis PSF, integrated over the full width at half maximum (FWHM) of the off-axis PSF. The latter is defined as the encircled energy within the FWHM of the off-axis PSF, normalized by the encircled energy without the VPM and LPM. In other words, the contrast compares the diffracted light from an on-axis point source to the signal from an off-axis companion of equal magnitude, whereas the throughput is a measure of the detected signal from the companion. Here, the azimuthal average of the on-axis PSF is used, but only off-axis PSFs with displacements along the $x$-axis are calculated for computational convenience. The algorithm is stopped when the contrast improvement between iterations is small, while also limiting the loss in throughput.

The performance of the VC with an LPM is shown in Fig. \ref{fig:offaxisperf} in terms of contrast and throughput of a companion imaged at an angular separation $\alpha$ from the star. The values achieved with the VC alone, EPMs, and no phase mask (E-ELT) are also shown for comparison. The mean contrasts and throughputs within 4-19 $\lambda/D$ are reported in Table \ref{table:1}. The E-ELT, without a coronagraph present, provides a mean contrast of $2.8\times10^{-4}$. This value is reduced to $3.6\times10^{-5}$ with VC2 and $4.9\times10^{-5}$ with VC4. The introduction of an LPM significantly reduces the contrast to the $10^{-7}-10^{-6}$ range for both of the optimization regions considered. The EPMs offer improvement over the VCs without LPMs, but yield contrast that is approximately 7--17 times greater than the VC+LPM combination.

\begin{table}
\caption{Mean contrast and throughput from $\alpha=4~\lambda/D$ to $\alpha=19~\lambda/D$.}             
\label{table:1}      
\centering                          
\begin{tabular}{c c c}        
\hline\hline                 
Design & Contrast & Throughput \\    
\hline                        
		E-ELT & $2.8\times10^{-4}$ & 1.00    \\
		VC2 & $3.6\times10^{-5}$  & 0.85    \\
		VC4 & $4.9\times10^{-5}$  & 0.75   \\
		VC2+LPM1 & $4.0\times10^{-7}$ & 0.46     \\
		VC2+LPM2 & $1.2\times10^{-6}$ & 0.27    \\
		VC4+LPM3 & $4.2\times10^{-7}$ & 0.43    \\
		VC4+LPM4 & $1.2\times10^{-6}$ & 0.26    \\
		EPM1 & $2.9\times10^{-6}$ & 0.48    \\
		EPM2 & $2.1\times10^{-5}$ & 0.26    \\
\hline                                   
\end{tabular}
\end{table}

The LPMs and EPMs presented here were designed to yield roughly the same throughput. Specifically, the throughputs are approximately 0.5 for the semi-annular dark region and 0.3 for the full annulus. Like many advanced coronagraphs, the VC+LPM combinations offer improved contrast performance at the cost of off-axis throughput. In this case, the throughput is mainly a function of the inner boundary of the optimization region, which has been made as small as possible. Specifically, the inner boundaries used for VC2 are $\alpha=3~\lambda/D$ and $\alpha=3.25~\lambda/D$, for LPM1 and LPM3, respectively. For VC4, $\alpha=3.5~\lambda/D$ is used for both LPM2 and LPM4. In the case of the EPMs, $\alpha=2.25~\lambda/D$ and $\alpha=3.5~\lambda/D$ are used for EPM1 and EPM2, respectively. We note that significantly better contrast may be achieved with the optimization region further from the star. In addition, the resulting region of optimum contrast is typically smaller than the pre-defined optimization region due to edge effects. 

In general, the best design of a high-contrast imaging instrument is chosen with several performance aspects in mind, including the contrast, size and shape of the discovery region, off-axis throughput, and sensitivity to chromatic effects. Each of these are addressed in the following discussion. Sensitivity to practical errors such as imperfect wavefront control, alignment, vibration, and partial resolution of the star are also important and are considered in the next section. 

The desired contrast and dark region shape depends on the observational goal; that is, a full annulus is beneficial for disk imaging and discovering new companions, whereas only a partial annulus is needed for characterization of a known object. We also note that a semi-annulus may be sufficient for $360^\circ$ high-contrast imaging using vector-phase elements and clever polarization tricks \citep[e.g.][]{Otten2014}. 

For the methods present here, extending the dark region within the central lobe of the on-axis PSF does not allow the $\sim\!\!10^{-6}$ contrast to be preserved with the VC+LPM designs. In these cases, the algorithm does not converge to a solution with small contrast and large throughput values. Thus, in addition to observational goals, the optimal size and shape of the dark region depends on the size and shape of the central lobe of the on-axis PSF, and ultimately the telescope aperture and the coronagraph instrument. We also find that, in comparison to the full annular case, the inner boundary of a partial annular dark region may be forced closer to the optical axis without significantly degrading off-axis throughput performance. 

The outer edge of the optimization region, on the other hand, will likely be matched to the control region of the adaptive optics system. Here, we optimize within a $20~\lambda/D$ radius, which is representative of current state-of-the-art adaptive optics. We note, however, that instruments developed for the future E-ELT may have larger control regions thanks to next-generation deformable mirrors. Though the outer boundary does not significantly affect the contrast achieved, increasing the outer radius generally leads to higher spatial frequency phase variations in the calculated phase masks. Ultimately, the allowable phase variations are limited by the manufacturing process, which will be investigated in future work.

\begin{figure}[t]
	\centering
	\includegraphics{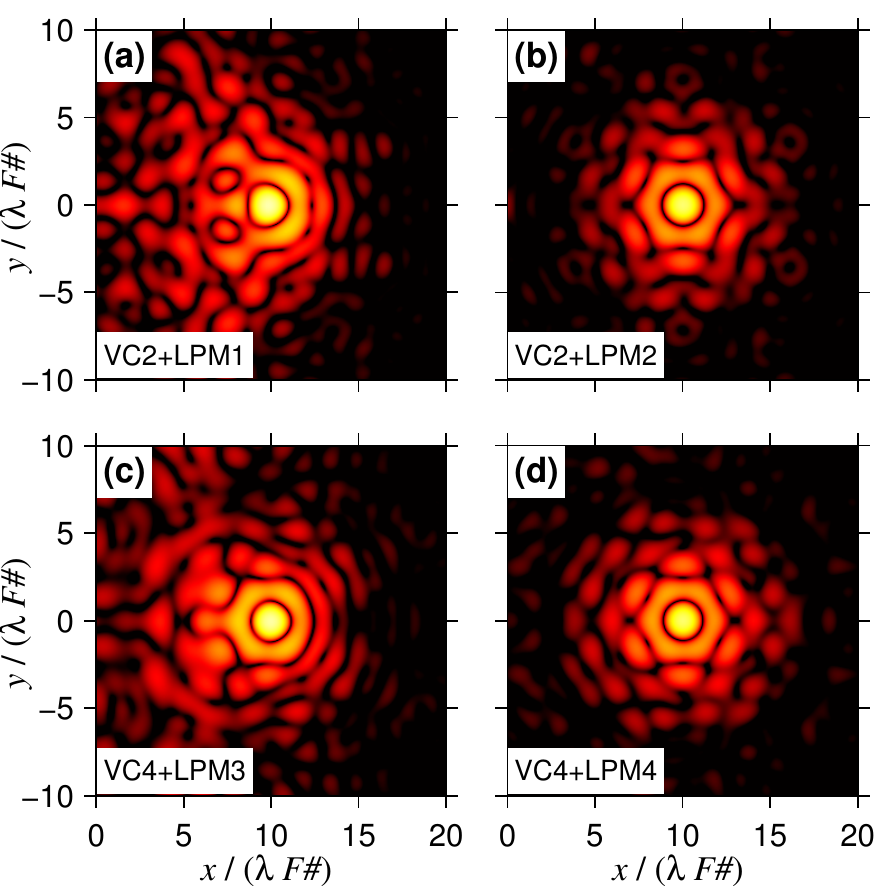}\\
	\includegraphics[trim = -5.5mm 0 0 0,clip=true]{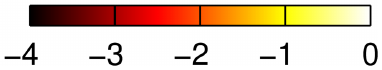}
	\caption{Off-axis PSF's (in log irradiance) along the $x$-axis for (a)~VC2+LPM1, (b)~VC2+LPM2, (c)~VC4+LPM3,  and (d)~VC4+LPM4 at $\alpha=10~\lambda/D$.}
	\label{fig:offaxisPSFs}
\end{figure}

An advantage of the VC+LPM, over other coronagraph designs such as those with occulting focal plane masks, is its intrinsic achromaticity. That is, assuming the phase masks apply perfectly achromatic phase shifts, the calculated power leaked through the LS, throughput, and contrast do not directly depend on wavelength. Rather, the wavelength dependence is limited to the scaling of the PSF, which causes radial blurring of the dark hole. Polychromatic light results in degraded contrast along a narrow annular strip at the edges of the optimization region where the affected width is given by roughly half the bandwidth fraction ($0.5~\Delta\lambda/\lambda$) multiplied by the inner and outer boundary angle. For example, with a bandwidth of $20\%$, the inner and outer boundaries of the dark hole are expanded and contracted by about $10\%$, respectively. Achromatic phase masks may be fabricated by direct writing of multilayer liquid crystals \citep{Komanduri2013,Miskiewicz2014,Otten2014}.

The width and shape of the off-axis PSF are very important for imaging applications. Figure \ref{fig:offaxisPSFs} shows the off-axis PSF for each LPM at $\alpha = 10~\lambda/D$. A relatively high-quality PSF is formed in each case (see Fig. \ref{fig:offaxisperf}(b),(d) for corresponding throughput values) and the shape is maintained for off-axis sources throughout the dark region of the on-axis PSF. Moreover, the PSF is spatially invariant over the optimization region, which is desirable for performing deconvolution on the obtained images as well as other forms of post-processing. 

\section{Sensitivity to errors}

A primary practical concern is that aberrations may cause starlight to appear inside the dark zone and, therefore, degrade contrast. In this section, the LPMs are shown to be robust to typical wavefront error expected for an adaptively-corrected high-contrast imaging instrument. The contrast is often limited by imperfections in the optical surfaces, which form quasi-static speckles in the image. Thus, we model the phase error in the pupil with the normalized power spectral density (PSD) function
\begin{equation}
PSD\left(\xi \right) = \left\{\begin{matrix}
1 & \xi \le \xi_0 \\
\left( \xi_0/\xi \right)^{2.5} & \xi > \xi_0\\
\end{matrix} \right. ,
\end{equation}
where $\xi$ is the magnitude of the spatial frequency and $\xi_0$ is the "cut-off" spatial frequency. In the following simulations, random phase screens are generated and scaled to root-mean-square wavefront error $\omega$. For reference, state-of-the-art systems typically have values $\xi_0=200~\mathrm{cycles/m}$ and $\omega \approx \lambda/100$.

\begin{figure}[t]
	\centering
	\includegraphics[width=\linewidth]{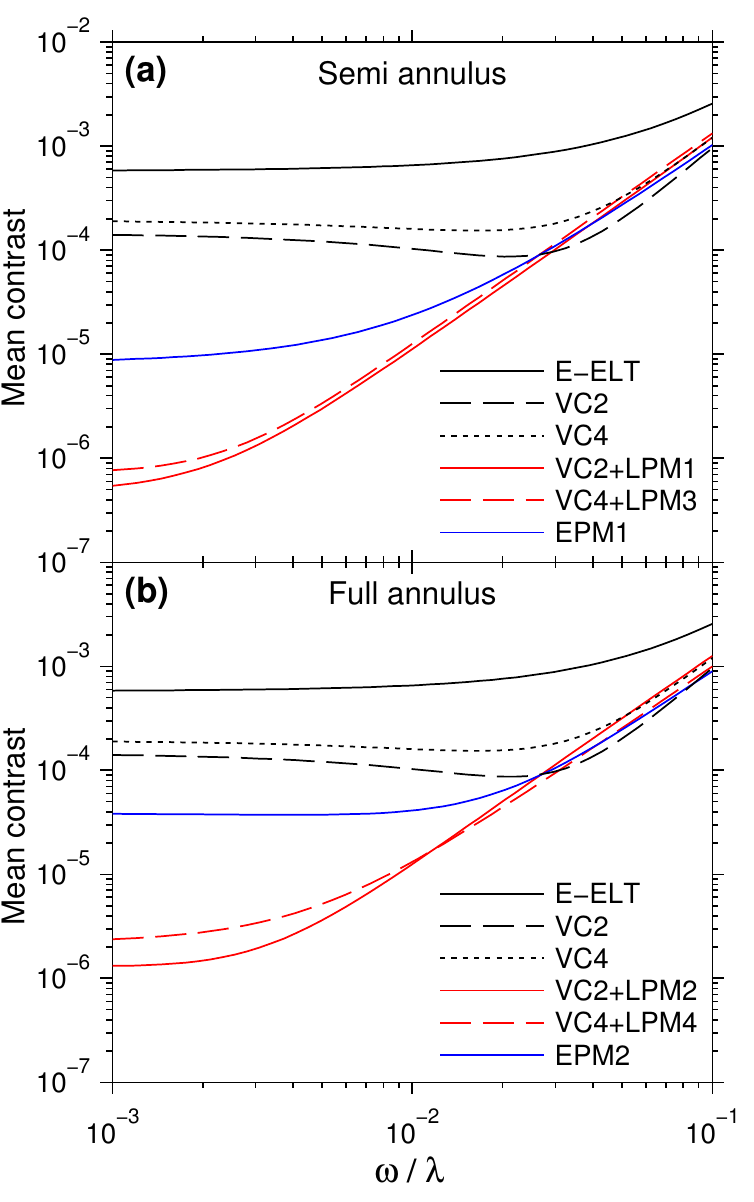}
	\caption{Mean contrast for a point source displaced along the $x$-axis (without azimuthal averaging) within $\alpha=4-19~\lambda/D$ as a function of root-mean-square phase error in PP1 $\omega$ for (a) semi-annular and (b) annular dark regions. }
	\label{fig:aberrations}
\end{figure}

To simulate realistic aberrations, a random phase screen is generated at PP1 and the resulting contrast is calculated. Figure \ref{fig:aberrations} shows the mean contrast for a point source displaced along the $x$-axis within $\alpha=4-19~\lambda/D$ as a function of $\omega$. Although unwanted speckles appear in the dark region, the VC and LPMs offer contrast improvement in the presence of wavefront errors potentially achieved on current and next-generation high-contrast imaging instruments ($\omega < \lambda/100$). Such aberrations levels require post-coronagraphic low-order wavefront sensing solutions, such as those proposed in \citet{Codona2013}, \citet{Singh2014,Singh2015}, or \citet{Huby2015}.

The sensitivity to low-order aberrations may also be controlled by the choice of focal plane mask. Specifically, the energy transmitted through the LS increases as $\alpha^{|l|}$ for $\alpha \ll \lambda/D$ and therefore an $l=4$ VPM may be more suitable than $l=2$ on a system susceptible to pointing errors and/or vibrations. In addition, reduced sensitivity to tip-tilt avoids leaked light owing to partial resolution of the star \citep[][]{Delacroix2014}. We note that the EPM designs are the least sensitive to these types of errors.

Imperfect alignment of the LPM will also negatively effect the achieved contrast performance. For the LPMs presented here, we expect the mean contrast to increase by a factor of $\sim\!\!1.5$ for an offset of 0.1\% of $D$. Alignment to within <0.2\% is possible in practice \citep{Montagnier2007}. 

\section{Conclusions}
We find that phase-only optical elements placed in the Lyot plane of a vortex coronagraph may improve the contrast performance on telescopes with complicated apertures. The combination of a vortex coronagraph and Lyot-plane phase mask provides better than $10^{-6}$ contrast within $\alpha=4-19~\lambda/D$ on heavily obscured telescopes, such as the E-ELT, at the expense of throughput. Moreover, we have shown that the improvement offered by an LPM is robust to realistic aberrations. Reducing the starlight, and its associated noise, enables sensitive high-contrast imaging of circumstellar disks and exoplanets. Phase masks similar to those described here are expected to improve the performance of current Lyot-style coronagraphs on ground-based telescopes, and may provide a route to terrestrial planet imaging with future space telescopes.

\begin{acknowledgements}
This work has benefited from fruitful discussions with Prof. Jean Surdej (Universit\'{e} de Li\`{e}ge, Belgium) and Prof. Matt Kenworthy (Leiden Observatory, Netherlands) as well as computing assistance from Carlos Gomez Gonzalez (Universit\'{e} de Li\`{e}ge, Belgium). G. J. R. was supported by Wallonie-Bruxelles International's (Belgium) Scholarship for Excellence and the U.S. National Science Foundation under Grant No. ECCS-1309517. The research leading to these results has received funding from the European Research Council under the European Union's Seventh Framework Programme (ERC Grant Agreement n. 337569) and from the French Community of Belgium through an ARC grant for Concerted Research Action. 
\end{acknowledgements}

\bibliographystyle{aa} 
\bibliography{Ruane_LPMs} 

\end{document}